# Magnetic Reconnection-Driven Energization of Protons up to ~400 keV at the Near-Sun Heliospheric Current Sheet

M. I. Desai[1,2], J. F. Drake[3,4], T. Phan[5], Z. Yin[3], M. Swisdak[3], D. J. McComas[6], S. D. Bale[5,7], A. Rahmati[5], D. Larson[5], W. H. Matthaeus[8], M. A. Dayeh[1,2], M. J. Starkey[1], N. E. Raouafi[9], D. G. Mitchell[9], C. M. S. Cohen[10], J. R. Szalay[6], J. Giacalone[11], M. E. Hill[9], E. R. Christian[12], N. A. Schwadron[13], R. L. McNutt Jr.[9], O. Malandraki[14], P. Whittlesey[5], R. Livi[5], and J. C. Kasper[15]

[1]Southwest Research Institute, 6220 Culebra Road, San Antonio, TX 78238, USA
[2]Department of Physics and Astronomy, University of Texas at San Antonio, San Antonio, TX 78249, USA
[3]Department of Physics, the Institute for Physical Science and Technology and the Joint Space Institute, University of Maryland, College Park, MD, USA.
[4]Institute for Research in Electronics and Applied Physics, University of Maryland, College Park, MD, USA
[5]Space Sciences Laboratory, University of California, Berkeley, CA
[6]Department of Astrophysical Sciences, Princeton University, NJ 08544, USA
[7]Physics Department, University of California, Berkeley, CA
[8]University of Delaware, Newark, DE 19716, USA
[9]Johns Hopkins University/Applied Physics Laboratory, Laurel, MD 20723, USA
[10]California Institute of Technology, Pasadena, CA 91125, USA
[11]The University of Arizona, Lunar and Planetary Laboratory, Tucson, AZ 85721, USA
[12]NASA Goddard Space Flight Center, Greenbelt, MD 20771, USA
[13]University of New Hampshire, 8 College Road, Durham NH 03824, USA
[14]Institute for Astronomy, Astrophysics, Space Applications & Remote Sensing (IAASARS), National Observatory of Athens, Greece
[15]University of Michigan, Ann Arbor, MI 48109, USA

**Abstract**

We report observations of direct evidence of energetic protons being accelerated above ~400 keV within the reconnection exhaust of a heliospheric current sheet (HCS) crossing by NASA's Parker Solar Probe (PSP) at a distance of ~16.25 solar radii ($R_s$) from the Sun. Inside the exhaust, both the reconnection-generated plasma jet and the accelerated protons up to ~400 keV propagated toward the Sun, unambiguously establishing their origin from HCS reconnection sites located anti-sunward of PSP. Within the core of the exhaust, PSP detected stably trapped energetic protons up to ~400 keV, which is ≈1000 times greater than the available magnetic energy per particle. The differential energy spectrum of the accelerated protons behaved as a pure power-law with spectral index of ~-5. Supporting simulations using the *kglobal* model suggest that the trapping and acceleration of protons up to ~400 keV in the reconnection exhaust is likely facilitated by merging magnetic islands with a guide field between ~0.2–0.3 of the reconnecting magnetic field, consistent with the observations. These new results, enabled by PSP's proximity to the Sun, demonstrate that magnetic reconnection in the HCS is a significant new source of energetic particles in the near-Sun solar wind. Our findings of in-situ particle acceleration via magnetic reconnection at the HCS provide valuable insights into this fundamental process which frequently converts the large magnetic field energy density in the near-Sun plasma environment and may be responsible for heating the sun's atmosphere, accelerating the solar wind, and energizing charged particles to extremely high energies in solar flares.

1. **Introduction**

Magnetic reconnection is the process in which magnetic field lines break and reconnect, converting magnetic energy into motional energy of charged particles in various natural and controlled environments, including terrestrial and planetary magnetospheres, solar and stellar flares, and fusion devices(Drake et al. 2006; Fleishman et al. 2022; Phan et al. 2018; Pontin & Priest 2001). In the heliospheric current sheet (HCS) where the interplanetary magnetic field (IMF) reverses its polarity, past spacecraft observations have shown that magnetic reconnection dissipates the Parker spiral magnetic field, generating high-speed flows and energizing particles(Eriksson et al. 2022; Gosling et al. 2006, 2005b; Lavraud et al. 2020; Phan et al. 2020; Szabo et al. 2020). However, there is vigorous debate about the extent of reconnection-associated particle energization in the HCS(Gosling et al. 2005a; Khabarova et al. 2020).

A key parameter that controls the heating and energization of electrons and ions during reconnection is the magnetic energy released per particle $m_i C_A^2$, where $C_A$ is the Alfvén velocity associated with the plasma parameters upstream of the reconnecting current layer(Phan et al. 2013, 2014; Shay et al. 2014; Øieroset et al. 2023). In reconnection events with direct evidence of strong particle energization, such as Earth's low-density magnetotail and solar flares, $m_i C_A^2$ is typically above 10 keV(Arnold et al. 2021; Øieroset et al. 2023). Conversely, the origin of the ~MeV energetic ions observed near HCS crossings around and beyond Earth orbit(Khabarova et al. 2015; Zank et al. 2014; Zharkova & Khabarova 2015) remains under debate since, at these distances, the corresponding $m_i C_A^2$ values are lower than 100 eV(Drake et al. 2009; Gosling et al. 2005a; Murtas et al. 2024). Recent observations from NASA's PSP have revealed unexpected ~20–100 keV suprathermal (ST) proton intensity enhancements during several near-Sun HCS crossings, despite having $m_i C_A^2$ values around 0.2 keV(Phan et al., 2022; Desai et al., 2022). However, since these energetic protons were found to propagate away from the Sun, it was unclear whether they originated from the Sun or from the HCS reconnection sites located sunward of PSP(Desai et al., 2022), thus leaving the possibility of strong particle energization during reconnection at the HCS as an open question.

Here we report direct evidence of reconnection at the near-Sun HCS accelerating protons up to ~400 keV. Since $m_i C_A^2$ for this event was only ~0.5 keV, surprisingly, the most energetic protons had energies nearly three orders of magnitude greater than the available magnetic energy per particle. A key component of these PSP data is the observation of a sunward-directed reconnection exhaust(Phan et al. 2024) and the concurrent detection of sunward-streaming energetic protons within the exhaust. Therefore, this event clearly demonstrates that energetic protons up to ~400 keV are accelerated at the HCS reconnection sites located anti-sunward of PSP and not by unrelated processes at the Sun.

## 2. Parker Solar Probe Observations

We use solar wind ion and electron data from SWEAP/SPAN(Kasper et al. 2016; Livi et al. 2022; Whittlesey et al. 2020), magnetic field data from FIELDS(Bale et al. 2016), and energetic proton measurements from IS☉IS(Hill et al. 2017; McComas et al. 2016). Figure 1 shows that the HCS crossing studied here occurred over a long duration (~3.7 hr) on December 12 from ~06:20 to ~10:00 UT during PSP solar encounter 14 (E14) at a distance of ~16.25 solar radii ($R_s$) from the Sun. The HCS crossing is recognized by the rotation of the radial component $B_R$ of the IMF from –650 nT to +470 nT (Figure 1h), the flipping of solar-origin strahl electron pitch angle (PA) fluxes from 180° to 0° (Figure 1e), and the factor of ~10 density enhancement (Figure 1f) relative to the solar wind, across the current sheet. Figure 1g shows that relative to the solar wind speeds outside the HCS, the plasma speed inside the HCS was smaller due to the persistent presence of a sunward-directed plasma jet emanating from a dominant reconnection site located anti-sunward of the spacecraft(Phan et al. 2024). The magnetic shear across the HCS was ~162°. Note that across the core of the exhaust $B_R<0$. This is a consequence of the differing Alfvén speeds on the two sides of the HCS. The magnetic field within a reconnection exhaust is tilted when the upstream Alfvén speeds differ(Lin & Lee 1993). The sketch in Figure 1i illustrates the approximate trajectory of the spacecraft through the HCS based on the variations of $B_R$. Since the outflow was continuously observed for 3.7 hours and the reconnection X-line driving this outflow was convecting outward with the ambient solar wind above 200 km s$^{-1}$, the inferred radial scale length of the exhaust as sampled by PSP was at least ~5 $R_s$.

The proton intensities between ~67 and 536 keV (Figure 1a) start increasing just before PSP enters the HCS at ~06:20 UT; those measured in the lowest energy range increase by nearly four orders of magnitude while those measured at higher energies show clear but somewhat smaller (between ~1-2 orders of magnitude) enhancements within the HCS. During the onset of the proton intensities (06:20-06:25 UT, Period 1 in Figure 1a), PSP traverses open field regions as evidenced from the anti-sunward strahl electrons seen near PAs of 180° in Figure 1e. Here, the energetic proton population flowing back toward the Sun shown in Figure 1b is not only more intense but also extends to higher energies up to ~400 keV when compared with that flowing away from the Sun shown in Figure 1c. The pitch angle distributions (PADs, Figure 1d) during the initial portion of the flat-top intensity enhancement show sunward-propagating energetic protons transitioning into near-isotropy followed by a peak near ~90° toward the end of Period 2. Just before the polarity reverses. i.e., throughout Period 3, PSP traverses a region where the PADs peak at 90°, indicating a trapped population of energetic protons where the nearly field-aligned and anti-field-aligned particles have escaped into the solar wind (see Figure 2 for details).

After the magnetic field reverses polarity at ~06:47 UT and until PSP exits the HCS at ~10:00 UT, the ~67-134 keV proton intensities decrease by approximately two orders of magnitude but are still ~100 times greater than those measured outside the HCS. The proton intensities exhibit at least two separate increases from ~0717-0805 UT and ~0806-0845 UT, labeled A and B in Figure 1a, of approximately the same ~30 to 40-minute duration as the enhancement observed prior to the polarity reversal (Figures 1a, 1b-1d). Throughout this post-polarity reversal interval, the differential intensities of the sunward-propagating protons with PAs between ~90°–180° are significantly greater (Figures 1b-1d) than those of the anti-sunward flowing protons with PAs between 0°–90°. Here, since $B_R$ points away from the Sun (Figure **1h**, $B_R>0$), most of these energetic protons originate from sources located anti-sunward of PSP and not the Sun. The intermittent presence of counterstreaming strahl electrons (Figure 1e) throughout the HCS encounter is another strong indicator that PSP is located sunward of the reconnection X-line and traversing reconnection-generated closed-field lines with both footpoints anchored at the Sun(Gosling et al. 2006). This is in contrast to the strahl dropouts that are typically observed when a spacecraft is located anti-sunward of the X-line and traverses reconnection-generated field lines disconnected from the Sun(Gosling et al. 2005b). Briefly, Figure 1 demonstrates that from ~06:25 UT until ~10:00 UT, PSP detected closed magnetic field lines produced by reconnection at the HCS and simultaneously observed a sunward-flowing plasma jet and energetic protons up to approximately 400 keV from HCS reconnection sources located anti-sunward of PSP.

Figure 2 examines the temporal evolution of the proton PADs and their energy dependence from 06:20–07:00 UT. During this 40-minute interval, PSP entered the HCS, and detected abrupt onsets and flat-top profiles in the proton intensities (Figure 1a). During the onset – Period 1 from ~06:20-06:25 UT (Figure 2d) – the proton PADs peak at ~120° PA and show an additional smaller peak near ~45°, possibly because protons near these two PAs are reflected back into the exhaust from external regions with enhanced field strength but those with PAs near 0°, 90°, and 180° are "lost" in magnetospheric-type loss-cones(Borovsky et al. 2022).

During the initial portion of the flat-top intensity enhancement from 06:26-06:37 UT (Period 2, Figure 2e), the proton PADs between ~90 and 400 keV exhibit strong (~10:1 anisotropy) sunward flow with broad peaks over 0°-45° PAs, while those between ~70 and 90 keV show peaks between 45° and 135° indicating the presence of a quasi-trapped proton population at these lower energies. Just before the magnetic field reverses polarity at 06:47 UT, the proton PADs at all energies up to 400 keV exhibit ~30°-45° wide peaks centered at ~90° (Period 3, Figure 2f). In this region, the field is relatively weak (~250 nT) with a low $B_R$ and a dominant $B_N$ component (Figures 2c and 1h), which implies that the ~67-400 keV protons with Larmor radii ranging between ~150-370 km (Hörandel 2010) are ExB drifting and efficiently trapped perpendicular to the local magnetic field, and further that PSP is traversing regions close to their acceleration sites near the core of the reconnection exhaust. After the magnetic field polarity reverses at ~06:47 UT, the proton PADs (Figure 2g) associated with the intensity modulations observed between ~06:48-10:00 UT in Figure 1a show that the sunward-streaming energetic proton population is nearly uniformly distributed over ~90°-180° PAs and extends upwards of ~400 keV. In summary, with the magnetic field pointing radially inward during Period 2 (Figures 2c, 2e) and outward during Period 4 (Figures. 2c, 2g), the PADs of protons up to ~400 keV during these two intervals exhibit strong sunward anisotropy, thereby providing direct, compelling evidence that these energetic protons originate from HCS reconnection sources located anti-sunward of PSP.

The differential intensity versus kinetic energy of protons from solar wind energies through the ST energy regime (Figure 3a) shows that the proton spectrum between ~67 and 527 keV behaves as a power law of the form $dj/dE \propto E^{-\gamma}$, where $dj/dE$ is the differential intensity at energy E in MeV and $\gamma \sim 5.1 \pm 0.12$ is the fitted value of the spectral index. Additionally, the HCS solar wind proton spectrum shows the formation of a high energy tail between 5 and 20 keV, which is well above the core Maxwellian distribution. Extrapolating the energetic proton spectrum to lower energies into the solar wind regime reveals the formation of a shoulder on the distribution, which might indicate the presence of local wave growth and scattering that interferes with the continuous acceleration to higher energies(Verniero et al. 2020; Fitzmaurice et al. 2024).

### 3. Simulations of Magnetic Reconnection

The magnetic energy released per particle controls the heating and energization of electrons and ions during reconnection. For a system with an asymmetry in the upstream conditions, this parameter is given by $m_i C_{Ah}^2$ where $C_{Ah}$ is the hybrid Alfvén velocity for asymmetric magnetic reconnection, given by $C_{Ah} = [B_1 B_2 (B_1+B_2)/4\pi m_p (n_1 B_2 + n_2 B_1)]^{1/2}$, where the subscripts 1 and 2 denote the magnetic field magnitude and density either side of the HCS(Cassak & Shay 2007; Øieroset et al. 2023; Phan et al. 2013, 2014; Shay et al. 2014). In this HCS event, $C_{Ah}$ was ~232 km s$^{-1}$, so $m_i C_{Ah}^2$ was ~0.56 keV. Thus, it is surprising that the protons form power-law distributions that extend upwards of 400 keV, which is ~$10^3$ times greater than the available magnetic energy per particle.

To establish that such extreme proton energization is possible in this event and to better understand the underlying physical mechanisms, we carried out simulations with the *kglobal* model(Arnold et al., 2019, 2021; Drake et al., 2019), which has been upgraded to describe the energization of ions as well as electrons(Yin et al. 2024a, 2024b). We carry out 2D simulations with a Harris-like equilibrium and upstream parameters based on direct PSP measurements outside of the reconnection exhaust. The anti-parallel component of the magnetic field (determined using the minimum variance analysis of the magnetic field (Sonnerup & Cahill Jr. 1967)) at the two edges of the HCS are given by $B_1$ = 560 nT and $B_2$ = 470 nT with corresponding densities of 1500/cm$^3$ and 3000/cm$^3$(Phan et al. 2024), respectively. The upstream proton and electron temperatures are 46 eV and 91 eV (not shown), respectively. The simulations are carried out with an initial current layer thickness of 0.01L and two different guide fields, corresponding to 0.2 and 0.3 of the reconnecting fields. Both these values are within the range of uncertainty in the guide field determination due to the long (3.7 hours) duration of the HCS crossing. The proton to electron mass ratio is 25 so the electrons are more massive than reality. However, it has been well established that the reconnection dynamics is insensitive to the electron mass(Hesse et al. 1999; Shay et al. 2007; Shay & Drake 1998). Magnetic reconnection is controlled by a hyper-resistivity as in earlier simulations with fluid ions(Arnold et al. 2021). The key characteristic of the model is that all kinetic scales have been eliminated so that direct simulation of reconnection dynamics in macroscale systems is now possible. All scales are normalized to the system scale length L, which is the periodicity length along the direction of the reconnecting magnetic field.

The simulations revealed that the HCS developed multiple reconnection sites resulting in the formation of large numbers of flux ropes that merged dynamically during the evolution of the system (see Figures 4a-4e). Reconnection proceeds with the growth of many small flux ropes that undergo mergers to drive electron and proton acceleration(Drake et al. 2006, 2013; Oka et al. 2010). The feedback of energetic particles on the reconnection dynamics occurs predominantly through the pressure anisotropy. Because particle heating and acceleration occur predominantly through Fermi reflection(Dahlin et al. 2014; Li et al. 2019) and the pressure-driven macroscale parallel electric field, the pressure parallel to the ambient magnetic field greatly exceeds that in the perpendicular direction. Figure 4f shows the firehose parameter $\alpha = 1 - 4\pi(P_\parallel - P_\perp)/B^2$ late in time from the simulation in Figure 4e. Negative values of α correspond to regions that are firehose unstable. The cores of islands are near marginal firehose stability so that the magnetic tension force that drives reconnection is strongly reduced. There, the plasma has undergone strong energization and the energetic particles feedback on the dynamics of the flux ropes to suppress reconnection and island merger(Arnold et al. 2021; Drake et al. 2006, 2013). The magnetic field $B_z$ increases substantially within the flux ropes compared with the corresponding upstream value due to plasma compression in the reconnecting current layer (Figure 4g). The enhanced values of $B_z$ reduce the magnetic shear during the flux rope merger compared with that across the HCS prior to reconnection(Phan et al. 2024).

### 4. Particle Acceleration in Merging Flux Ropes

Figure 4h shows the initial and late time distributions of electrons and protons, with the protons reaching ~500 keV for the simulation with a guide field of 0.2. While the protons contain modestly more energy than the electrons at late times, simulations show that both species are strongly energized. The strong energization of both electrons and protons results from the formation and merging of magnetic flux ropes in the reconnecting current layers. Electrons initially have a higher upstream temperature than the protons, but the protons reach higher energies up to ~500 keV compared with the ~100 keV energy attained by the electrons because of efficient Fermi reflection during the flux rope merger.

The power-law spectral indices of the energetic particles during reconnection is largely determined by the strength of the ambient guide field(Arnold et al. 2021). This is because an increasing guide field increases the effective radius of curvature of the reconnecting magnetic field thereby reducing the effectiveness of the Fermi drive mechanism(Arnold et al. 2021; Dahlin et al. 2016; Drake et al. 2006). Figure 3b shows the early (Figure 4a) and late time (Figure 4e) distributions of the proton particle flux for two values, namely 0.2 and 0.3, of the ambient guide magnetic field. The simulation with a guide field of 0.3 provides the best match to the observed proton power-law spectral index of approximately -5 shown in Figure 3a. The simulation with a guide field of 0.2 has a slightly harder spectrum with a power-law index of approximately -4 with maximum proton energy of ~500 keV while that from the larger guide field is ~200 keV. This reflects the weaker acceleration of particles during reconnection with a stronger guide field. In *kglobal* simulations, the upper energy limit of the power-law distribution is controlled by the size of the domain(Arnold et al. 2021; Yin et al. 2024a), thus a larger domain with a guide field of 0.3 is also anticipated to yield 500 keV protons. Briefly, the *kglobal* simulations reveal that protons up to ~500 keV are produced in a spectrum of merging flux ropes (Figures 4a-4e) during reconnection even when the available magnetic energy-per-particle is only approximately 0.5 keV. That a small number of protons can gain such significant energy is a consequence of the rate of energy gain during Fermi reflection being proportional to the energy of the particles(Drake et al., 2006, 2013; Guo et al., 2014; Arnold et al., 2021; Zhang et al., 2021). Once a subset of particles gains sufficient energy, this population can rapidly outrun the bulk of the protons and gain significantly more energy than the average available per particle as reconnection proceeds.

5. **Summary and Discussion**

We have presented five key observational features that show that mechanisms associated with magnetic reconnection in the near-Sun HCS energized solar wind protons into extended power-laws up to energies ≈1000 times greater than the available magnetic energy per particle. These observations are: 1) the presence of a sunward-directed plasma jet from reconnection sources located anti-sunward of PSP(Phan et al. 2024); 2) the intensities of sunward-flowing protons up to ~400 keV are greater than that of the anti-sunward-flowing population throughout the ~3.7-hr HCS crossing, indicating that the energetic particle source regions are also located anti-sunward of PSP; 3) near the core of the HCS reconnection exhaust, the proton PADs up to ~400 keV peak at ~90°, indicating the presence of a stably trapped population of energetic protons as PSP traverses regions very close to their acceleration sites; 4) intermittent presence of counterstreaming electron strahls indicate that reconnection processes occurring in the HCS beyond PSP orbit produced closed magnetic field lines with footpoints connected to the Sun(Gosling et al. 2006); and 5) the core solar wind proton distribution develops a shoulder that likely extends into the energetic proton regime upwards of ~400 keV, with the ~67-527 keV proton spectrum behaving as a power-law with a spectral index of ~5.1.

Observations and computer simulations of charged-particle acceleration during reconnection reveal that the three primary mechanisms responsible are: Fermi acceleration in Alfvénic outflows and island mergers; betatron acceleration or magnetic pumping; and direct acceleration by the magnetic field-aligned electric field(Birn et al. 2012; Dahlin et al. 2014; Drake et al. 2006; Guo et al. 2019; Li et al. 2019; Oka et al. 2022). While wave-particle interactions and resonant acceleration may also be present, their contributions are expected to be less significant(Ergun et al., 2020a). Reconnection-generated turbulence or reconnection in a turbulent medium can also contribute to the acceleration processes in multiple stages(Ergun et al. 2020a; Lazarian et al. 2020; Matthaeus & Lamkin 1986). Indeed, observations in the Earth's magnetotail and supporting test-particle simulations indicate that reconnection-generated turbulence can create large-amplitude electric fields near the ion cyclotron frequency with local magnetic field depletions that can trap and accelerate electrons and ions to produce power-law distributions(Ergun et al., 2020a,b). In contrast, throughout the ~3.7-hr HCS crossing (Figures 1h, 2c) and especially when PSP detected the ≲400 keV trapped proton population in Period 3, the magnetic field remained relatively steady at ~200 nT and exhibited few signatures of turbulence similar to those seen in the magnetotail(Ergun et al., 2020b). Instead, for this HCS encounter, large variations in the normal component of the magnetic field ($B_N$ in Figures 1h, 2c) indicate that large-scale magnetic islands or flux ropes are embedded within the exhaust. In addition, the identification of several subscale current sheets within the HCS reconnection exhaust points to merging flux ropes(Phan et al. 2024) that most likely play an important role in trapping and accelerating the energetic particles(Drake et al. 2006, 2013, 2019; Li et al. 2017; Oka et al. 2010; Zank et al. 2014).

We explored particle energization during this reconnection event using the *kglobal* model(Yin et al. 2024b). The proton distributions formed extended power laws with energies reaching approximately 500 keV and 200 keV with power-law slopes of -4 and -5, for the guide fields of 0.2 and 0.3, respectively. These simulations suggest that the trapping and acceleration of protons into power-laws up to ≈400 keV in the near-Sun HCS reconnection exhaust is likely facilitated by merging magnetic islands (see Phan et al. 2024) with a guide field around ~0.25 of the reconnecting magnetic field. These simulations reproduced the ~1000-fold energization of protons and generated a power-law spectral index of around -5, which aligned well with the observations.

Finally, our results have significant implications for particle energization in solar flares. It has long been hypothesized that nonthermal electrons and protons in flares carry comparable energy(Emslie et al. 2012). However, direct evidence for energetic protons in flares comes from observations of gamma-ray flares, which are relatively uncommon and only provide information on protons with energy above ~1 MeV(Lin et al. 2003). Additionally, critical information about proton energies in flares below ~1 MeV is lacking(Vogt & Henoux 1999), even though in-situ heliospheric measurements of various ion species following impulsive flare events reveal power-law spectra that extend down to ~100's of keV(e.g., see Mason 2007). Reconnection events in the Earth's magnetotail reveal power-law spectra of both electrons and protons(Oieroset et al., 2002; Ergun et al. 2020b). However, the proton power-laws extend only a single decade in energy, possibly due to demagnetization effects in the narrower current layers that characterize the magnetotail reconnection events. In this HCS reconnection event, however, the available magnetic energy per particle, $m_i C_A^2$, is only ~0.56 keV, while the proton power-law extended above ~400 keV, nearly 1000 times $m_i C_A^2$. In a typical flare event, with B~50 G and n~5x10$^9$ cm$^3$, the parameter $m_i C_A^2$ is around 10 keV. Thus, by analogy, we can expect solar flares to accelerate protons from 10's of keV up to ~10 MeV. In summary, these definitive measurements of in-situ particle energization in the near-Sun HCS provide fresh insights into magnetic reconnection processes that routinely dissipate the large magnetic field energy density in the near-Sun plasma environment and may be responsible for accelerating charged particles to near-relativistic speeds in solar flares(Fleishman et al. 2022; Pontin & Priest 2001), heating the sun's atmosphere, and driving the solar wind(Bale et al. 2023).

**Acknowledgements**

This work was supported by NASA's Parker Solar Probe Mission, contract NNN06AA01C. We thank all the scientists and engineers who have worked hard to make PSP a successful mission, particularly the engineers, scientists, and administrators who designed and built the ISOIS/EPI-Lo, ISOIS/EPI-Hi, FIELDS, and SWEAP instrument suites, and who supported their operations and the scientific analysis of the data. For their contributions to the scientific configuration and instrumental analysis, we owe special thanks to P. Kollmann, J. Peachy, and J. Vandegriff at JHU/APL for EPI-Lo. PSP data are available at the NASA Space

Physics Data Facility. Work at SwRI is supported in part by NASA grants 80NSSC20K1815, 80NSSC18K1446, 80NSSC21K0112, 80NSSC20K1255, and 80NSSC21K0971. W. H. M. is partially supported by the PSP/IS☉IS project at the University of Delaware subcontract SUB0000317 from Princeton University and by NASA PSP Guest Investigator grant 80NSSC21K1765.

Figures and Figure Captions

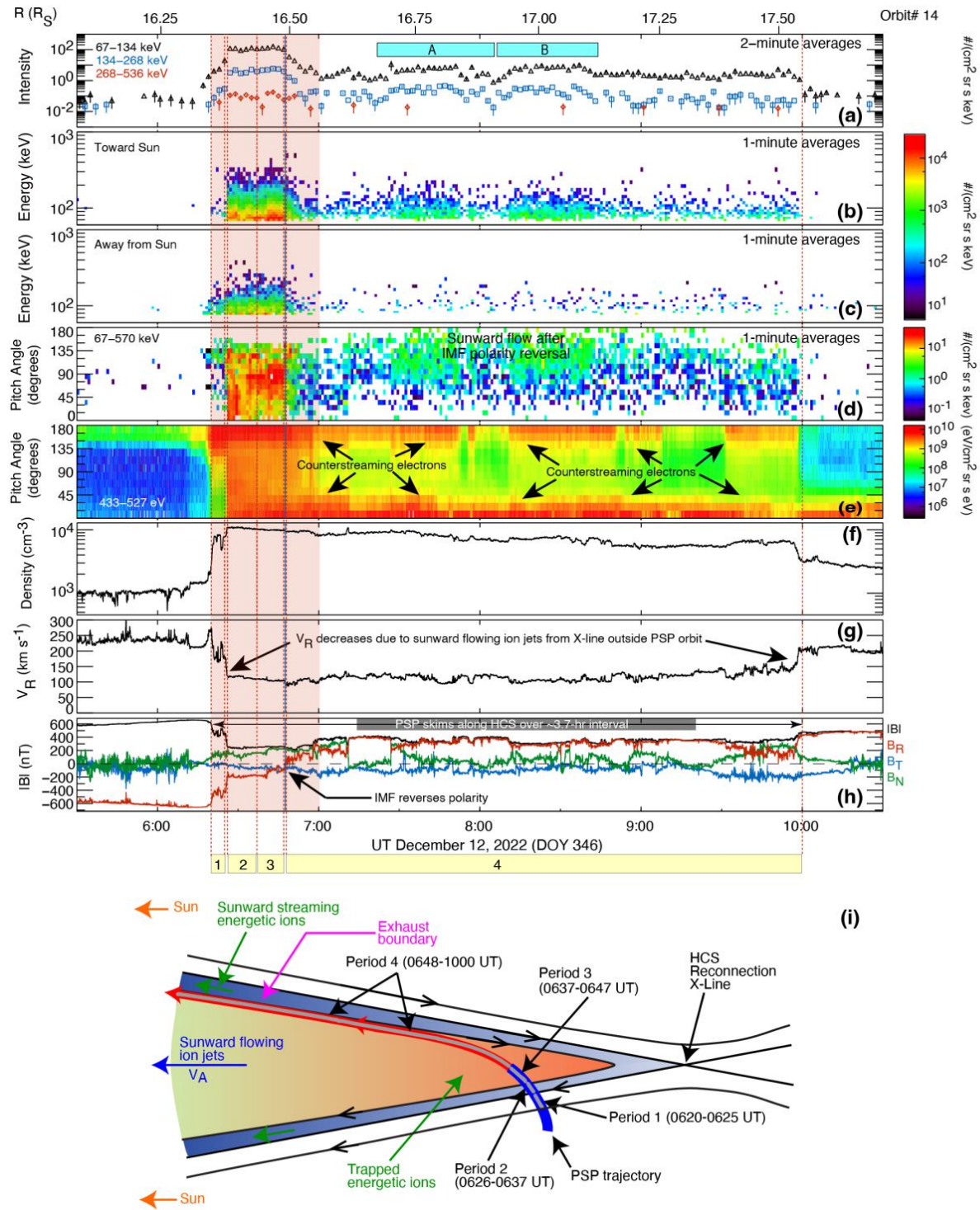

**Figure 1:** Overview of energetic proton(McComas et al. 2016), solar wind electron(Whittlesey et al. 2020), solar wind ion(Livi et al. 2022), and magnetic field(Bale et al. 2016) observations and approximate Parker Solar Probe (PSP) trajectory associated with the heliospheric current sheet (HCS) crossing during solar encounter 14 (E14). Time profiles from 05:30–10:30 UT on December 12, 2022 of: (a) 2-minute omni-directional averages (over 80 EPI-Lo apertures) of ST proton intensities in three energy ranges between ~67-536 keV; energy spectrograms of ST proton intensities integrated over (b) the anti-sunward-facing EPI-Lo apertures, defined as those with look directions between 120°-180° from the Sun-PSP vector, and (c) the sunward-facing apertures with look directions within 0°-60° of the Sun-PSP vector; (d) pitch-angle distributions (PADs), where pitch angle is the angle between the particle velocity and the magnetic field vector, of ~67-570 keV protons transformed into the solar wind frame using the Compton-Getting technique(Compton & Getting 1935; Ipavich 1974; Zhang 2005) to eliminate convective anisotropies resulting from the bulk solar wind flow; (e) energy flux spectrogram of 433-527 eV electrons which constitute the magnetic field-aligned electron beam or strahl originating directly from the solar corona(Phan et al. 2020); (f) proton density measured by SWEAP/SPAN-Ions (see (Phan et al. 2024); (g) radial component of the solar wind velocity; and (g) magnetic field magnitude, and its vector components in the sun-centered RTN coordinate system (radial:red, tangential:blue, and normal:green). The red vertical lines show the ~3.7 hr duration of the E14 HCS crossing and the four intervals identified at the bottom that are selected for detailed analysis in Figure 2. The blue vertical line denotes the time of the IMF polarity reversal, i.e., when the radial component $B_R$ in Figure 1h changes sign from negative or inward to positive or outward. (i) A two-dimensional sketch showing the approximate PSP trajectory projected on the plane of reconnection. The blue line along the trajectory depicts time intervals and corresponding regions where PSP observed negative or inward IMF polarity and the red portion shows time intervals and regions of positive or outward polarity.

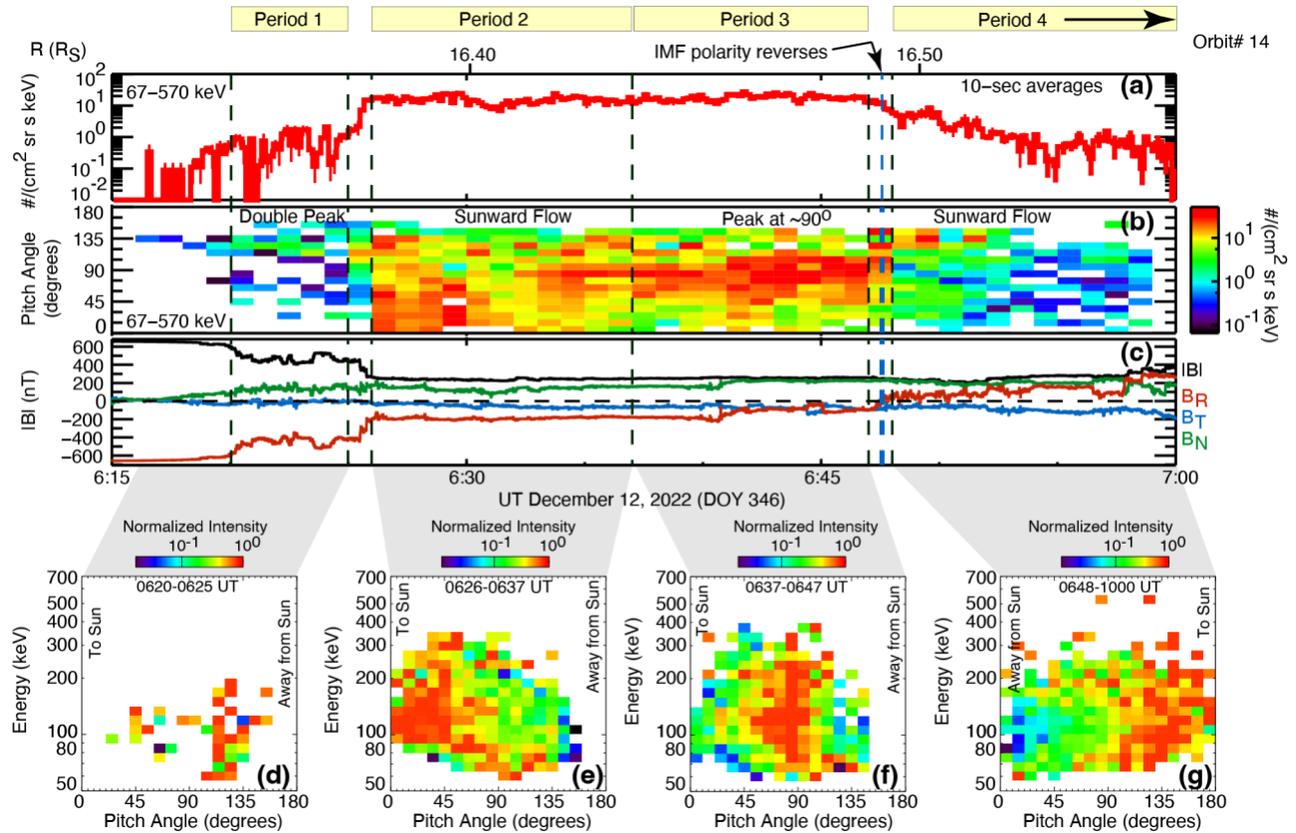

**Figure 2.** Evolution of pitch-angle distributions during the PSP E14 HCS crossing. (a) High resolution (10-seconds) omni-directional averages of the intensities, and (b) SW frame PADs of ~67-570 keV protons; (c) magnetic field magnitude and vector components from 06:15-07:00 UT on Dec. 12, 2022. (d)-(g) Color-coded normalized differential intensities plotted as a function of energy versus pitch angle during periods 1-4 identified in Figure 1 and Figures 2a-2c. During each period, the intensity in each energy bin is normalized to the maximum intensity observed at that energy. The lack of measurements at ~155°-180° PAs during Periods 1-3 is due to the relatively sparse or complete lack of coverage of these PAs by EPI-Lo.

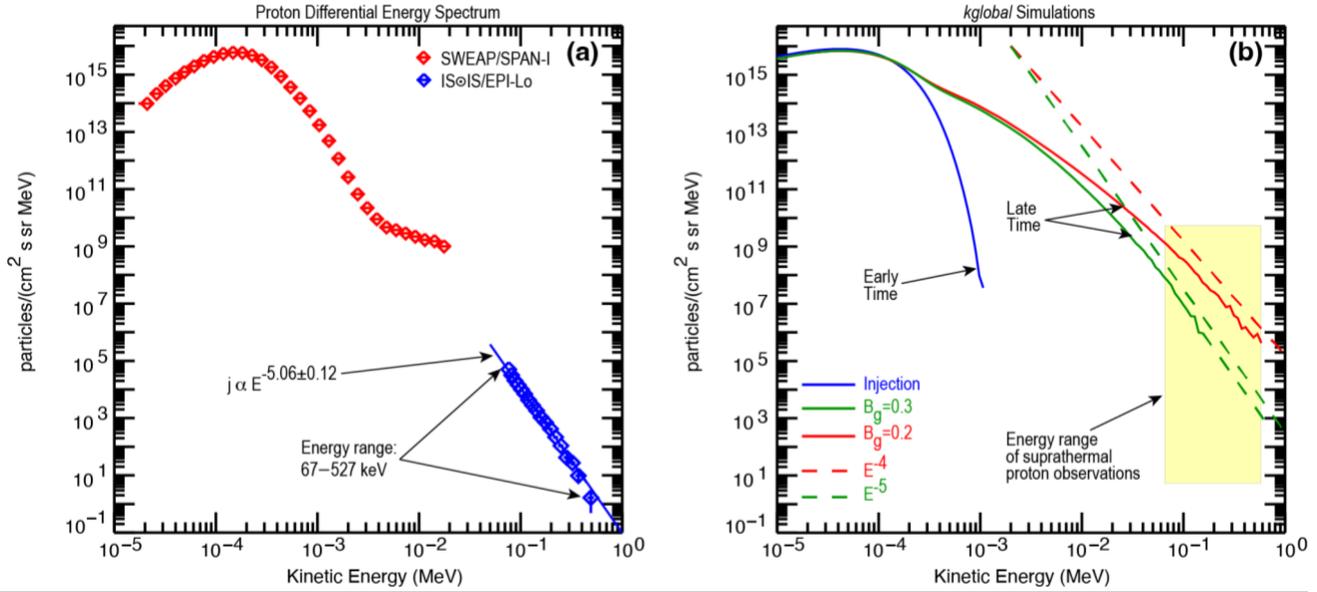

**Figure 3:** Proton differential energy spectrum during the E14 HCS crossing and comparison with *kglobal* simulations. (a) Differential energy spectrum of ~200 eV–527 keV protons from 06:20-10:00 UT. The solid line shows a power-law fit of the form dJ/dE **α** $E^{-\gamma}$ to the proton spectrum between 67 and 527 keV. The spectral index γ along with the measured proton energy range is shown. Note that this proton spectrum exhibits a notably harder or flatter profile and reaches energies nearly five times greater than that of the 30-100 keV proton spectrum (with γ values around 8) recorded by PSP in association with fast solar wind microstreams, which may be generated by interchange reconnection processes occurring near the Sun(Bale et al. 2023). (b) Proton spectra from *kglobal* simulations(Arnold et al. 2019; Drake et al. 2019; Yin et al. 2024a) of reconnection at the HCS for two representative guide fields of 0.2 and 0.3 of the reconnecting magnetic field. Early and late times in the simulations are described in Figure 4.

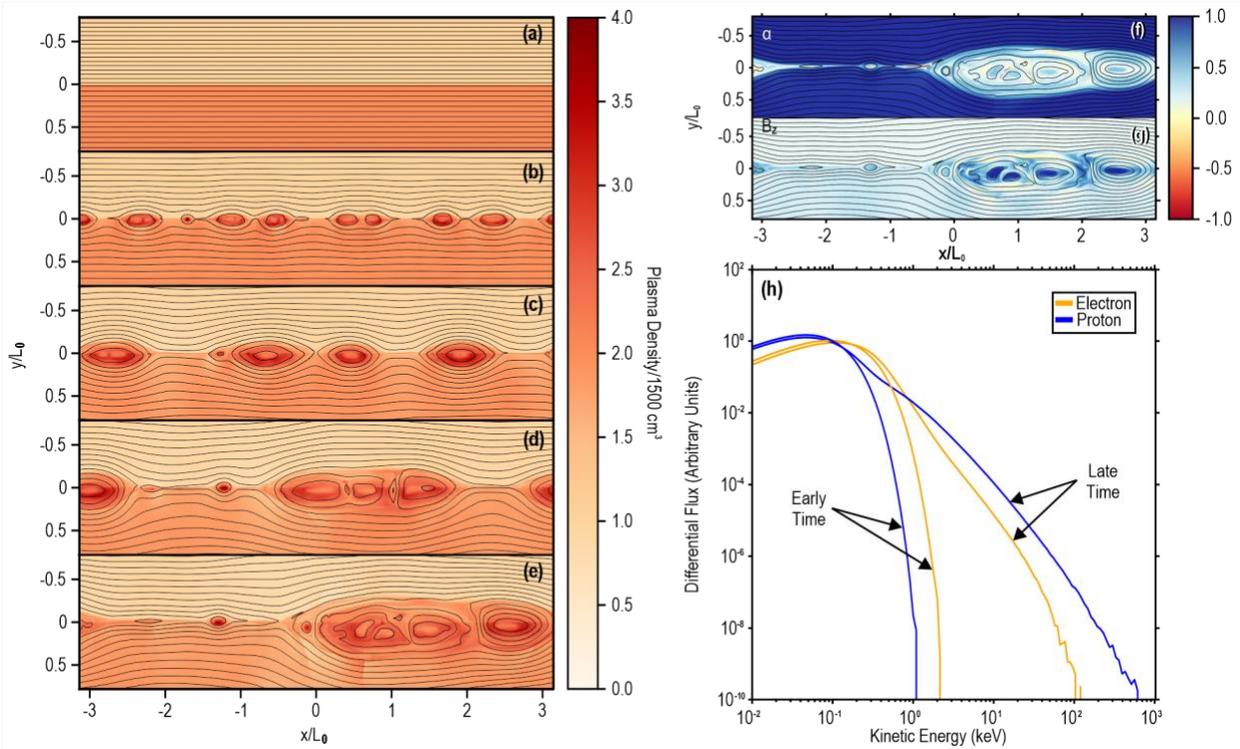

**Figure 4:** Left: Time development of flux ropes in a simulation of the E14 HCS reconnection event. From top (a) to bottom (e), current sheet evolution at five times, $t/\tau_A$ = 0, 4, 8, 12, and 16, where $\tau_A = L_0/C_{A0}$ is the Alfvén transit time across the simulation domain from *kglobal* simulations with an ambient guide field of 0.2 times the upstream magnetic field. The color bar indicates the plasma density normalized to 1500/cm$^3$. The black lines are the in-plane magnetic field. Early time is defined as $t/\tau_A$ = 0, and late time is $t/\tau_A$ = 16. Right: Figures 4f and 4g respectively show the firehose parameter $\alpha = 1 - 4\pi(P_\parallel - P_\perp)/B^2$ and the out-of-plane magnetic field $B_z$ late in time from the simulation in Figure 4e. The color bar represents $\alpha$ and $B_z$ normalized to the upstream magnetic field value. Right: Figure 4h shows the differential flux of electrons and protons at early and late times from the simulation of Figures 4a-4e.